\documentclass{article}
\usepackage[utf8]{inputenc}
\usepackage{arxiv}
\usepackage{amsmath}
\usepackage{enumerate}
\usepackage{graphicx}
\usepackage{booktabs}
\usepackage{url}
\usepackage[numbers]{natbib}

\graphicspath{{.}}

\usepackage{rotating}
\usepackage{caption}
\usepackage{subcaption}
\usepackage[inline,shortlabels]{enumitem}

\usepackage{amsmath}
\usepackage{amssymb}
\usepackage{mathtools}
\usepackage{algorithm,algorithmic}

\usepackage[nameinlink]{cleveref}
\crefname{figure}{Fig.}{Figs.}
\Crefname{figure}{Figure}{Figures}
\crefname{equation}{Eq.}{Eqs.}
\Crefname{equation}{Equation}{Equations}
\crefformat{section}{#2§#1#3}
\crefname{section}{§}{§§}
\Crefname{section}{Section}{Sections}
\crefname{table}{Table}{Tables}
\crefname{appendix}{Appendix}{Appendices}

\DeclareMathOperator*{\E}{\mathbb{E}}
\DeclareMathOperator{\softmax}{softmax}

\title{Distill2Explain:Differentiable decision trees for explainable reinforcement learning in energy application controllers}

\author{%
Gargya Gokhale~\thanks{Under review}\\
IDLab, Ghent University--imec\\
Gent, Belgium \\
\texttt{gargya.gokhale@ugent.be} \\
\And
Seyed Soroush Karimi Madahi\\
IDLab, Ghent University--imec\\
Gent, Belgium\\
\And
Bert Claessens\\
IDLab, Ghent University--imec\\
beebop.ai \\
\And
Chris Develder\\
IDLab, Ghent University--imec\\
Gent, Belgium\\
}

\begin{document}

\maketitle

\begin{abstract}
Demand-side flexibility is gaining importance as a crucial element in the energy transition process. Accounting for about 25\% of final energy consumption globally, the residential sector is an 
important (potential) source of energy flexibility. 
However, unlocking this flexibility requires developing a control framework that (1)~easily scales across different houses, (2)~is easy to maintain, and (3)~is simple to understand for end-users. A potential control framework for such a task is data-driven control, specifically model-free reinforcement learning (RL). Such RL-based controllers learn a good control policy by interacting with their environment, learning purely based on data and with minimal human intervention. Yet, they lack explainability, which hampers user acceptance. Moreover, limited hardware capabilities of residential assets forms a hurdle (e.g., using deep neural networks). To overcome both those challenges, we propose a novel method to obtain explainable RL policies by using differentiable decision trees. Using a policy distillation approach, we train these differentiable decision trees to mimic standard RL-based controllers, leading to a decision tree-based control policy that is data-driven and easy to explain. As a proof-of-concept, we examine the performance and explainability of our proposed approach in a battery-based home energy management system to reduce energy costs. For this use case, we show that our proposed approach can outperform baseline rule-based policies by about 20-25\%, while providing simple, explainable control policies. We further compare these explainable policies with standard RL policies and examine the performance trade-offs associated with this increased explainability.
\end{abstract}


\section{Introduction}
\label{sec:intro}
The ongoing shift towards sustainable energy is leading to a significant restructuring of the energy sector: large-scale integration of distributed renewable energy sources, increased electrification, phasing out of fossil fuel-based generation, etc.~\cite{irena}. As a result of these changes, there is a growing need for grid balancing services and demand-side flexibility to ensure reliable and secure functioning of the grid. Conventionally, large industries and big consumers were the primary source of such demand-side flexibility. However, another important and as-of-yet untapped source of flexibility is the residential sector~\cite{iea-2023}. 

Households account for about 25\% of the final energy consumption and with growing adoption of rooftop solar PVs, home batteries, heat pumps, etc., represent an appealing source of flexibility~\cite{energy-cons-houses}. Usually, exploiting this flexibility entails optimizing the use of a battery or other flexible assets to shift the real-time consumption of households while ensuring user comfort~\cite{energy-flex-buildings}. In most cases, the primary objective is to minimize the energy bill of the household, however, prior research has also investigated other objectives such as maximizing self-consumption or participation in other explicit demand response services~\cite{hems-review-1, hems-review-challenges}. 

An important component for extracting this household flexibility is a home energy management system~(HEMS), responsible for solving the underlying, non-linear sequential decision-making problem and calculating the necessary control actions to be taken in real-time. Developing HEMS has been a major research area, with works such as~\cite{hems-review-3, hems-review-2} providing an overview of techniques used in literature.

A prominent research direction in this context is the use of model-predictive control (MPC) algorithms. MPC forms an advanced control framework that relies on a model of the system to predict the system’s behavior and uses the model to analytically obtain optimal actions~\cite{mpc-basic}. Works such as~\cite{hems-mpc-e-energy, real-mpc, mpc-hems-2} have investigated the application of MPC in both simulation and real-world scenarios, showing significant performance improvements in such systems. However, as highlighted in~\cite{mpc-problems-2, mpc-problems-1}, accurate models---which an MPC requires---of the system are often difficult to obtain in the residential sector, significantly limiting widespread adoption of MPC-based solutions in this sector. 

The residential sector necessitates control frameworks that can easily scale to many, potentially diverse households. This has led to an increased interest in data-driven control frameworks, especially based on reinforcement learning (RL). RL-based controllers work by continuously interacting with the environment (i.e., the household), collecting experience (data) from these interactions, and using them to learn a control policy that maximizes a predefined reward~\cite{rl-basic}. Thus, with little human intervention and relying completely on data, such RL-based controllers can learn good control policies. Previous works on RL-based HEMS controllers such as~\cite{rl-hems-1, safe-rl-hems-1, rl-hems-2} have shown significant improvements over baseline scenarios. 

However, most RL-based research is limited to simulation environments or specialized buildings. As discussed in~\cite{nagy2023ten}, this is due to two main factors: 
 \begin{enumerate*}[(i)]
  \item \label{it:data-inefficiency} the data inefficiency of RL training, and 
  \item \label{it:opaque} the opaque nature of obtained control policies
\end{enumerate*}.
To address \ref{it:data-inefficiency}, i.e., the high amount of data required for training RL-based controllers, previous works such as~\cite{hybrid-mpc, meta-rl-hems, transfer-rl} propose different solutions. 
However, \ref{it:opaque}~raises another important concern related to RL, i.e., the non-interpretable/non-explainable nature of their policies, especially when based on deep neural networks. With limited prior works in this area, we identify this as a significant gap in existing literature and thus introduce our innovative approach to specifically address the (lack of) explainability of RL-based HEMS. 

More specifically, we propose a policy distillation framework using differentiable decision trees~\cite{soft-dt-2, soft-dt-1}. The key idea is to distill information from pre-trained RL-based controllers into an explainable decision tree, leading to control policies that are explainable and perform nearly as good as the original RL-based policies. To the best of our knowledge, this is one of the first works in the energy field to adopt policy distillation using differentiable decision trees for explainable RL.
Our main contributions can be summarized as:
\begin{enumerate}[noitemsep,nolistsep]
    \item We propose a novel framework for explainable RL that uses differentiable decision trees and policy distillation for converting black-box RL policies into explainable decision trees~(\cref{sec:method}).
    \item Using different case studies, we detail the explainability of our proposed method, contrasting it with conventional RL-based policies~(\cref{subsec:exp_eval}).
    \item We compare the performance of our method with conventional RL-based policies and other baselines to show the performance trade-off that results from the increased explainability~(\cref{subsec:performance_eval}). 
\end{enumerate}
The primary emphasis of this paper is to introduce a novel method for obtaining explainable RL-based control policies. As a  proof-of-concept, we validate our proposed approach on a battery-based home energy management scenario using real-world data and present our preliminary findings. Section~\cref{subsec:future} outlines the future work in terms of additional investigation of this method and its application to other, more complex scenarios.



\section{Related Work}
\label{sec:related}
Designing control algorithms for unlocking flexibility in households has been a major field of research, with works such as~\cite{hems-review-3, hems-review-2} providing an exhaustive review of prior works including heuristics-based controllers, MPCs, and data-driven algorithms. As discussed in~\cref{sec:intro}, our work focuses on improving the explainability of reinforcement learning-based controllers and hence this section focuses on developments in the fields of reinforcement learning-based control, policy distillation, and explainable AI. We refer interested readers to~\cite{hems-review-3, hems-review-challenges} for more comprehensive reviews of other relevant methods in the context of HEMS and demand-side flexibility. 
\subsection{Data-driven Home Energy Management Systems}
A recent research direction in HEMS has been the use of data-driven and mainly reinforcement learning-based controllers~\cite{rl-review}. RL-based controllers rely primarily on past data and have minimal modeling requirements as compared to prominent control techniques such as MPCs. For example, works such as~\cite{rl-hems-1, safe-rl-hems-1, rl-hems-2}, demonstrate the applications of RL-based controllers in the context of HEMS. In most of these cases, the RL-based controllers rely on state-of-the-art RL algorithms such as deep $Q$-networks (DQN)~\cite{dqn}, deep deterministic policy gradient (DDPG)~\cite{ddpg} and 
use control policies based on deep neural networks to achieve significant performance improvements ($\sim$5-16\% as reported in these works). While these deep neural networks are beneficial for achieving good performance, a common drawback associated with their use is their opaque (or black-box) control policy~\cite{rl-challenges}. 
We aim to address this challenge associated with the explainability of RL-based controllers, providing a framework for distilling a standard RL control policy into an explainable policy. 

\subsection{Explainable AI}
Providing explainability for AI-based technology is an important and necessary issue 
to address for large-scale deployment of machine learning-based solutions, especially in the context of the energy sector. 
We refer to more exhaustive reviews~\cite{xai-review-1, xai-review-2} of available techniques, metrics, and methodologies across different fields such as image recognition, natural language processing, etc. However, as discussed in~\cite{xai-energy-review}, in the context of energy, research 
on explainable AI has been largely restricted to applications such as forecasting, modeling, or fault diagnosis. While few works such as~\cite{xai-hems-rl-decom, xai-multiagent-rl, xai-shap-rl}, present explainable RL-based controllers, they largely rely on decomposition methods or utilize post-hoc explanation frameworks such as SHAPley values, feature importances, LIME, etc. Although useful, such post-hoc explanations are typically designed for experts and are not easily accessible to the average end-user, such as a homeowner. Our proposed method 
differs from such approaches by distilling the deep RL-based control policy into an explainable architecture in the form of differentiable decision trees. 
Thus, the resulting control policies are structurally explainable, i.e., 
in the form of rather simple \emph{if-then-else} rules, 
that can be easily
\begin{enumerate*}[(a)]
\item explained to non-expert end users, and
\item deployed 
on simple hardware.
\end{enumerate*}

\subsection{Policy Distillation and Differentiable Decision Trees}
As discussed in~\cref{sec:intro}, our approach employs policy distillation to trained RL-based controllers and distills their knowledge into a differentiable decision tree structure. This closely follows prior works that have used knowledge distillation strategies to 
\begin{enumerate*}[(1)]
    \item compress large neural networks, or
    \item combine knowledge from model ensembles into a single model~\cite{distillation-review, knowledge-distill-hinton}.
\end{enumerate*}
Differing from these, works such as~\cite{rl-distillation-ddt, rl-distillation-fuzzy, rl-distillation} 
adopt knowledge distillation in RL to transform the architecture of the final policy, e.g., into a fuzzy inference system. We follow a similar approach and distill an RL-based control policy into a differentiable decision tree. This enables us to extract knowledge from standard RL-based controllers into simple decision trees which are structurally easy-to-explain and simple to understand. This choice is closely related to the objective of obtaining control policies that are easy to explain (to both energy experts and end-users). 

Differentiable decision trees~(DDTs) or soft decision trees are variants of binary decision trees that can be trained using gradient descent~\cite{soft-dt-2, soft-dt-1}. Prior works such as~\cite{knowledge-distill-hinton, ddt-regression} have applied DDTs to 
computer vision and regression tasks.
For our energy use case,
we follow the approach 
of~\cite{rl-distillation-ddt}, 
using DDTs to distill RL-based control policies.
However, our proposed approach differs from~\cite{rl-distillation-ddt} in two ways:
\begin{enumerate*}[(i)]
    \item we learn deterministic decision trees instead of the soft decision trees, and 
    \item we learn using observed, explainable features 
    (as opposed to rather indirect pixel-based learning in~\cite{rl-distillation-ddt}).
\end{enumerate*}
This enables us to learn DDTs using gradient descent and then convert them into simple decision decision trees for inference (as detailed further in~\cref{sec:method}).



\section{Preliminaries}
The proposed differentiable decision tree-based policy distillation framework is examined in the context of a home energy management scenario. This section describes the problem formulation for this proof-of-concept and introduces basic concepts related to reinforcement learning~(RL). 

\subsection{Problem Formulation}
\label{subsec:problem}
In the context of home energy management, we consider an average Belgian household with a rooftop solar PV installation~(with generated power $P^{\text{pv}}_{t}$), non-flexible electrical load~($P^{\text{con}}_{t}$), and a home battery. We assume that this household is exposed to varying BELPEX day-ahead prices~($\lambda^{\text{con}}_{t}$) and a capacity tariff based on peak power. This leads to a joint optimization problem, where the HEMS must minimize the daily cost of both the energy consumption~($c^{\text{eng}}_{t}$) and the peak power~($c^{p}_{t}$). This optimization problem is modeled as:
\begin{subequations}
\begin{align}
\min_{u_{1}, \ldots u_{T}} & \sum_{t=1}^{T} c^{\text{eng}}_{t} + c^{p}_{t} \label{subeq:obj_fun} \\
\text{s.t.:} \ c^{\text{eng}}_{t} &= \begin{cases}
                                \lambda^{\text{con}}_{t} \ P^{\text{agg}}_{t} \ \Delta t &: P^{\text{agg}}_{t} \geq 0      \\
                                \lambda^{\text{inj}}_{t} \ P^{\text{agg}}_{t} \ \Delta t &: P^{\text{agg}}_{t} < 0\\
                                \end{cases} \quad \forall t \label{subeq:eng_cost} \\
                c^{p}_{t} &= \lambda^{\text{cap}} \ \max (P^{\text{agg}}_{t}, \ P^{agg}_{\text{min}})  \label{subeq:cap_cost} \\
                P^{\text{agg}}_{t} &= P^{\text{con}}_{t} + P^{\text{pv}}_{t} + u_{t} \qquad \qquad \ \forall t \label{subeq:u_phys}\\
                E_{t+1} &= \begin{cases}
                                E_{t} + \eta \ u_{t} \ \Delta t &: u_{t} \geq 0      \\
                                E_{t} + \frac{1}{\eta} \ u_{t} \ \Delta t &: u_{t} < 0\\
                            \end{cases} \quad \forall t \label{subeq:bat_mod} \\
                0 \leq E_{t} &\leq E^{\text{max}}; \
                u^{\text{min}} \leq u_{t} \leq u^{\text{max}}  \quad \forall t.
\end{align}
\label{eq:opt}
\end{subequations}

The battery is modeled using a linear model~(\cref{subeq:bat_mod} with charging/discharging actions $u_{t}$ and current energy level~($E_{t}$). The cost of energy consumed~($c^{\text{eng}}_{t}$) depends on the power consumed~($P^{\text{agg}}_{t}$) and the current injection and consumption prices~($\lambda^{\text{inj}}_{t}$ and $\lambda^{\text{con}}_{t}$ respectively). Similarly, the capacity cost~($c^{p}_{t}$) depends on the actual power consumed and the minimum power capacity contracted~\cite{capacity-tariff}. Furthermore, we assume $T = 24~\text{hours}$ and a time resolution $\Delta t = 1~\text{hour}$. 

The above-mentioned problem illustrates a real-world scenario that is pertinent in the present day where a household's HEMS needs to efficiently leverage the home battery to reduce the energy bill, taking charging/discharging actions dependent on the real-time price, solar PV production, and daily load consumption patterns. Accordingly, we further assume that the HEMS can only take discrete actions (a total of 5 related to 2 charging modes, 2 discharging modes, and 1 `do nothing’ mode). Nonetheless, our method can be easily extended to other action spaces as well. 

\subsection{Markov Decision Process}
We model the sequential decision-making problem presented in~\cref{subsec:problem} as a Markov Decision Process (MDP)~\cite{rl-basic}. The states~($\textbf{x}_{t} \in \mathbf{X}$) consist of the current price~($\lambda^{\text{con}}_{t}$), battery state-of-charge, non-flexible demand~($P^{\text{con}}_{t}$), and solar PV generation~($P^{\text{pv}}_{t}$). The actions~($u_{t} \in \mathbf{U}$) are the charging/discharging signals given to the battery. As stated above, we assume a discrete action space of 5 elements~(i.e., $\mathbf{U} = \{ -1, -0.5, 0, 0.5, 1\}$), with the possibility of extending it reserved for future work. The reward function~($\rho: \mathbf{X} \times \mathbf{U} \rightarrow \mathbb{R}$) is defined as the cost incurred for each time step $t$ and is modeled based on~\cref{subeq:eng_cost}, \cref{subeq:cap_cost}. The transition function~($f$) models the dynamics of the household, taking into account the (controllable) behavior of the battery along with (uncontrollable) real-time solar PV generation, and non-flexible power consumption. 

\subsection{Reinforcement Learning}
In RL, the goal of an agent is to find a policy $\pi: \mathbf{X}\rightarrow \mathbf{U}$ that minimizes the expected $T$-step cost~($J^{\pi}$) starting from an initial state $\mathbf{x}_{0} \in \mathbf{X}$ (\cref{eq:t-step-cost}).
\begin{equation}
    J^{\pi} = \sum_{t=0}^{T}\rho(\mathbf{x}_{t}, \pi(\mathbf{x}_{t}), \omega)
\label{eq:t-step-cost}
\end{equation}
This expected cost~$J^{\pi}$ can be expressed as a recursive function using a state-action value function, called $Q$-function:
\begin{equation}
    Q^{\pi}(\mathbf{x}_{t}, \mathbf{u}_{t}) = \E_{\omega} [\rho(\mathbf{x}_{t}, \mathbf{u}_{t}, \omega) + \gamma Q^{\pi}(\mathbf{x}_{t+1}, \pi(\mathbf{x}_{t+1}))] .
\label{eq:basic-q-function}
\end{equation}
Here, $\omega$ represents the stochasticity in the transition function~($f$) and can be attributed to exogenous factors. The discount factor is represented as $\gamma$.

For our work, we focus on the deep-$Q$ network~(DQN) algorithm~\cite{dqn}, where the $Q$-function is iteratively estimated using a deep neural network as a function approximator. The neural network-based $Q$-function (parameterized as $\hat{Q}_{\theta}$) is trained on a batch of data~($\mathcal{F}$) with the following loss term:
\begin{equation}
    \mathcal{L} = \E\left[\left(\hat{Q}_{\theta} (\textbf{x}_{t}, \textbf{u}_{t}) - \left( c_{t} + \min_{\textbf{u} \in \mathbf{U}} \hat{Q}_{\theta^{-}}(\textbf{x}_{t+1}, \textbf{u}) \right) \right)^2 \right]
\label{eq:dqn_loss},
\end{equation}
where, $c_{t} = \rho(\textbf{x}_{t}, u_{t}, \omega)$ is the observed cost value during the state transition from $\textbf{x}_{t}$ to $\textbf{x}_{t+1}$ and the expectation is over all elements of $\mathcal{F}$. For more details related to the DQN algorithm, we refer the readers to~\cite{dqn}. Note that our proposed method is agnostic to the choice of the RL algorithm and can be easily extended to other RL algorithms as well. 



\section{Methodology}
\label{sec:method}
This section details our proposed approach. We first mathematically formulate the differentiable decision tree architecture, followed by the policy distillation process. 

\subsection{Differentiable Decision Trees~(DDTs)}
Differentiable decision trees or soft decision trees are a variant of ordinary decision trees, introduced in prior works such as~\cite{soft-dt-1, soft-dt-2}. We follow the work presented in~\cite{ddt-rl}, where a DDT is formulated as a directed, acyclic graph consisting of nodes and edges. There are two types of nodes in a DDT: (1) decision nodes, characterized by a feature selection weights~($\beta$) and a threshold~($\phi$); and (2) leaf nodes containing a weight vector~($\mathbf{w}^{L}$) to express the probability distribution. While ordinary decision trees have decision nodes represented using a boolean function, DDTs implement a `soft' decision using the \emph{sigmoid} function~(represented as $\sigma$). Consequently, each path (or edge) going out of a decision node carries a probability value that is based on the condition evaluated at that decision. 

\subsubsection{Decision Node}
\begin{figure}[t]
    \centering
    \includegraphics[width=0.7 \textwidth]{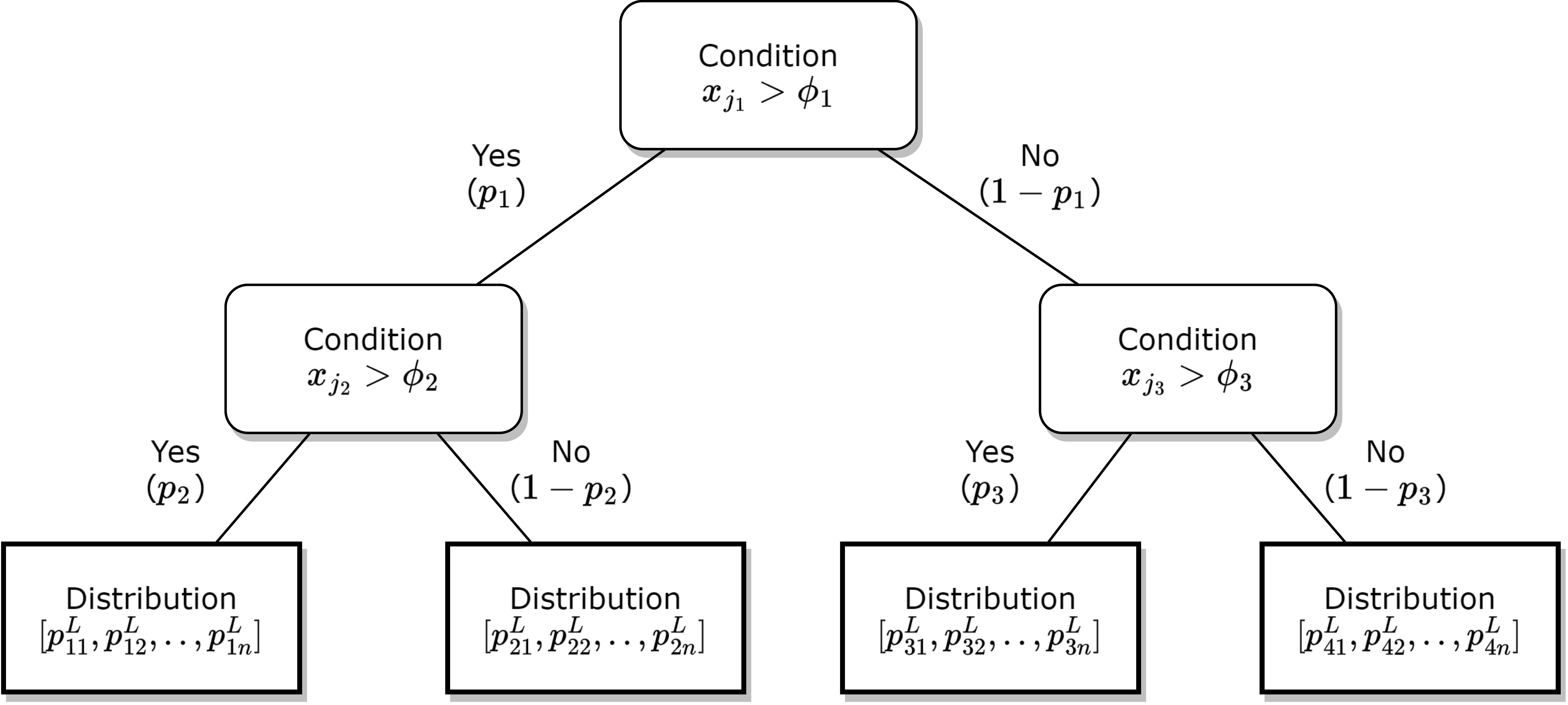}
    \caption{Illustration of a DDT of depth 2. The rounded boxes depict the decision nodes and the rectangles depict leaf nodes. All $p_{i}$ represent the path probabilities and $p^{L}_{jk}$ denotes the leaf probability distributions.}
    \label{fig:depth2_ddt}
\end{figure}

A decision node (represented as rounded boxes in~\cref{fig:depth2_ddt}) is modeled as:
\begin{subequations}    
    \begin{align}
        p &= \sigma \ (\mathbf{\beta} \mathbf{x} - \phi)     \\ 
        p^{\text{left}} &= p             \\
        p^{\text{right}} &= 1-p
    \end{align}
\label{eq:decision_node}
\end{subequations}

Here, $\mathbf{\beta}$ and $\phi$ are trainable parameters representing the feature selection weight and the cut thresholds respectively. Each decision node evaluates a condition based on the selected feature and cut threshold and gives path probabilities for going left (the condition is likely to be True) and going right (the condition is likely to be False). 

\subsubsection{Leaf Nodes}
A leaf node $l$ contains a weight vector~($\mathbf{w}^{L}_{l}$) that leads to an output probability distribution modeled using a \emph{SoftMax} function. In our case, each leaf output is the probability distribution over all actions in the action space~($\mathbf{U}$), however, this can be extended to estimate exact values~(for continuous actions) as well. For this leaf node, the probability for each action $u_{m} \in \mathbf{U}$ is calculated using~\cref{eq:leaf_prob}
\begin{equation}
    p^{L}_{lm} = \frac{e ^ {-w_{m}}}{\sum_{\kappa=1}^{|\mathbf{U}|} e ^ {-w_{\kappa}}} \ \ \forall m \in \{1, 2, \ldots, |\mathbf{U}|\}
\label{eq:leaf_prob}
\end{equation}

\subsubsection{Creating a DDT}
\cref{eq:decision_node} and \cref{eq:leaf_prob} are combined to implement a DDT of required depth. To illustrate this, we now present the formulation of a DDT of depth 2 (as shown in~\cref{fig:depth2_ddt}). Such a DDT contains 3 decision nodes and 4 leaf nodes. For each decision node, we have feature selection vectors~($\beta_{1}$, $\beta_{2}$, $\beta_{3}$) and cut-thresholds~($\phi_{1}$, $\phi_{2}$, $\phi_{3}$); each leaf node contains weight vectors~($\mathbf{w}^{L}_{k}$). The tree is built based on~\cref{alg:ddt_d2}. This formulation is used to perform a forward pass of the DDT and train the parameters using gradient descent. At inference, each node is converted from the `soft’ version into a crisp node, resembling an ordinary decision tree. This includes reducing all feature selection parameters~($\mathbf{\beta}$) into one-hot representations~(using \emph{argmax}) and converting all probabilities into `crisp', boolean values. Note that this method of creating a DDT decomposes all computations into differentiable computations and allows to parallelize them. Additionally, while a DDT of any depth can be implemented based on~\cref{eq:decision_node}, \cref{eq:leaf_prob}, for this work we restrict the scope to trees of depth 2 and 3. This choice is primarily driven by the ease of explainability for such (shallow) trees. 

\begin{algorithm}[t]
\begin{algorithmic}[1]
\STATE Initialize: $\boldsymbol{{\beta}_{i}}$, $\boldsymbol{\phi}$, $\textbf{w}_{k}^{L}$, where $i = \{1, 2, 3\}$ (decision nodes)  and $k = \{1, 2, 3, 4\}$ (leaf nodes)
\STATE Input: State $\textbf{x}$
\FORALL{i}
    \STATE Feature Selection: $x_{j} = \boldsymbol{{\beta}_{i}} \cdot \textbf{x}  $
    \STATE Evaluate Condition: $p_{i} = \sigma (x_j - {\phi}_{i})$
\ENDFOR
\STATE Calculating Path Probabilities: $ \textbf{p} = 
                            \begin{bsmallmatrix}
                                p_{1} &  0\\
                                0 &  1-p_{1}
                            \end{bsmallmatrix} 
                            \cdot
                            \begin{bsmallmatrix}
                                p_{2} &  1 - p_{2}\\
                                p_{3} &  1-p_{3}
                            \end{bsmallmatrix}
                            $
\FORALL{k}
    \STATE Calculate Leaf Probabilities: $\mathbf{p}^{L}_{k} = \{ p^{L}_{k1}, p^{L}_{k2}, \ldots p^{L}_{kn} \}$ based on~\cref{eq:leaf_prob}
\ENDFOR
\STATE Output: $o = \textbf{p}[1, 1] \textbf{p}^{L}_{1} + \textbf{p}[1, 2] \textbf{p}^{L}_{2} + \textbf{p}[2, 1] \textbf{p}^{L}_{3} +\textbf{p}[2, 2] \textbf{p}^{L}_{4} $
\end{algorithmic}
\caption{Depth 2 DDT Formulation}
\label{alg:ddt_d2}
\end{algorithm}

\subsection{Policy Distillation}
Distillation is a method for transferring knowledge from a teacher model $T$ to a student model $S$~\cite{rl-distillation}. In the context of reinforcement learning, this refers to transferring knowledge related to a control policy from a trained teacher agent~($\pi^{T}$) to a student agent~($\pi^{S}$). Typically, this leads to a classification problem where targets are obtained using the outputs of the trained agent. 

We follow the approach presented in~\cite{rl-distillation}, where a DQN-based teacher agent is trained first and then using a batch of observations~($\mathcal{F}$), a student policy is distilled based on the teacher agent. First, the trained teacher agent is used to create a new batch of training data of the form $\mathcal{D} = \{\mathbf{x}_{i}, \textbf{q}_{i}\}_{i=1}^{|\mathcal{F}|}$. Here, $\mathbf{q}_{i}$ is the vector corresponding to $Q$-values for all actions for a state~$\textbf{x}_{i} \in \mathcal{F}$, obtained using the teacher agent (i.e., $\mathbf{q}_{i} = \{ Q^{T}(\mathbf{x}_{i}, u_i) \mid \forall u_{i} \in \mathbf{U}\} $). Following this, the student agent is trained to mimic this distribution using Kullback-Leibler~(KL) divergence with temperature~($\tau$) as presented in~\cref{eq:student_loss}.
\begin{equation}
    \mathcal{L}_{\theta_{s}} = \softmax\left(\frac{\textbf{q}_{i}}{\tau}\right)
    \> \cdot \> \ln \left(\frac{\softmax\left(\frac{\textbf{q}_{i}}{\tau}\right)}{\softmax\left(\frac{\textbf{q}^{S}_{i}}{\tau}\right)}\right)
\label{eq:student_loss}
\end{equation}
Note that $\textbf{q}^{S}_{i}$ is the output of the student model parameterized by $\theta_{s}$. The temperature $\tau$ is used to adjust the `smoothness' of the $Q$-function distribution. 

\subsection{Our Approach}
For our work, we assume a teacher agent~(policy $\pi^{T}$ and $Q$-function $Q^{T}$) as a standard DQN agent, and the student agent~($\pi^{S}$) consists of the DDT architecture. First, the teacher agent is trained independently using DQN, to obtain a control policy. Following this, the trained teacher is used to create target distributions using data collected from previous interactions with the environment. This data is then used to train the student DDT-based agent. \Cref{alg:our_algo} outlines the training procedure for our proposed approach. 

\begin{algorithm}[t]
\begin{algorithmic}[1]
\STATE Initialize: Teacher agent $T$, DDT student $S$, buffer $\mathcal{F}$.
\STATE Train Teacher:\\
        Use $\mathcal{F}$ and \cref{eq:dqn_loss} to train teacher i.e., obtain $\pi^{T}$ and $Q^{T}$
\STATE Create Distillation Batch: \\
        Distillation batch $\mathcal{D} = \{\textbf{x}_{i}, \textbf{q}_{i}\}_{i=1}^{|\mathcal{F}}|$\\
        where $\mathbf{q}_{i} = \{ Q^{T}(\mathbf{x}_{i}, u_i) \mid \forall u_{i} \in \mathbf{U}\}$
\STATE Train Student DDT:\\
        Use $\mathcal{D}$ and \cref{eq:student_loss} to train the student~($\pi^{S}$) using gradient descent.
\end{algorithmic}
\caption{Training algorithm for our proposed method}
\label{alg:our_algo}
\end{algorithm}



\section{Experiment Setup}
We validate our proposed approach on a home energy management scenario using a battery as the source of flexibility. This section presents the simulation environment and details the training and experimental scenarios used. 

\subsection{Simulator Setup}
\label{subsec:simulator}
We use a Python-based simulation environment to validate and compare our proposed approach with standard RL-based controllers. This simulator is derived from a real-world Belgian household with rooftop solar PV and is modeled based on~\cref{eq:opt}. Real demand and solar PV profiles are used along with the battery model presented in~\cref{subeq:bat_mod}. Additionally, we use hourly, real-world BELPEX prices as consumption prices~($\lambda^{\text{con}}_{t}$) and a capacity tariff structure based on~\cite{capacity-tariff}. The battery parameters are detailed in~\cref{subsec:battery_hp}. Further, we assume the injection price~($\lambda^{\text{inj}}_{t}$) is $25\%$ of the consumption price i.e.,~($\lambda^{\text{inj}}_{t}= 0.25 \ \lambda^{\text{con}}_{t}$).

\subsection{Training Setup}
The training is divided into two parts:
 \begin{enumerate*}[(i)]
  \item training the teacher agent; and
  \item policy distillation to train student agent
\end{enumerate*}. 
For the teacher agent, we follow the standard DQN implementation and use an $\epsilon$-greedy training strategy to train the DQN-based teacher agent. Following this, the buffer generated by the DQN-based teacher is used to create the distillation dataset~($\mathcal{D}$). The student agent is then trained using this dataset. To improve the stability of the training process, we set the temperature~($\tau$) from~\cref{eq:student_loss} to 0.03 to obtain a sharp Q-function distribution. We list all the hyperparameters used in~\cref{subsec:dqn_hp}. For each agent (teacher and student) we perform 5 seeded runs and compare the mean values over the 5 runs. 

\subsection{Experiment Scenarios}
\label{subsec:scenarios}
The primary goal of this work is to present a novel approach for obtaining explainable, RL-based policies. We investigate the HEMS scenario described in~\cref{subsec:problem}, where one intentionally simplified scenario is used to effectively assess the explainability of our proposed approach. We specifically investigate two key scenarios: 

\subsubsection{Scenario 1: Performance Comparison}
In this scenario, we investigate the performance of our proposed approach and evaluate whether our method can achieve satisfactory performance compared to standard DQN agents and baseline rule-based controller. For this, we consider a realistic HEMS scenario and use real-world data for load profiles, solar PV, and prices as described in~\cref{subsec:simulator}. The performance is quantified as the total cost for a day, comprising both energy and capacity costs. As baselines, we use the teacher agents as the upper bound of performance and a rule-based control~(RBC) policy as the lower bound. The RBC policy is designed similar to the typical built-in control policy of home batteries and aims to maximize self-consumption. We consider two key variants of price profiles:
\begin{enumerate*}[(i)]
    \item an artificial, square wave price profile~(resembling day-night tariff); and
    \item an actual real-world day ahead price profile
\end{enumerate*}. 
The artificial price profile is a simplified scenario with clear peaks and valleys in the price to provide unambiguous opportunities for energy arbitrage. 

\subsubsection {Scenario 2: Explainability Assessment}
To further assess the explainability of our method, we consider a simplified scenario where we exclude solar PV from the system and reduce the state features to 3 components i.e., battery state-of-charge, price, and demand. This simplification enables clear visualization of the learned policies, contrasting them with standard DQN policies to qualitatively investigate the explainability of our proposed method.\footnote{Quantitatively assessing the explainability of AI methods remains an open question with most prior works relying on either qualitative methods or user studies for assessment~\cite{qunat-xai}.} 


\section{Results}
\label{sec:results}
This section presents the results obtained for the different scenarios discussed in~\cref{subsec:scenarios}. 
\subsection{Performance Evaluation}
\label{subsec:performance_eval}
\begin{figure}[t]
     \centering
     \begin{subfigure}[b]{0.45\textwidth}
         \centering
         \includegraphics[width=\textwidth]{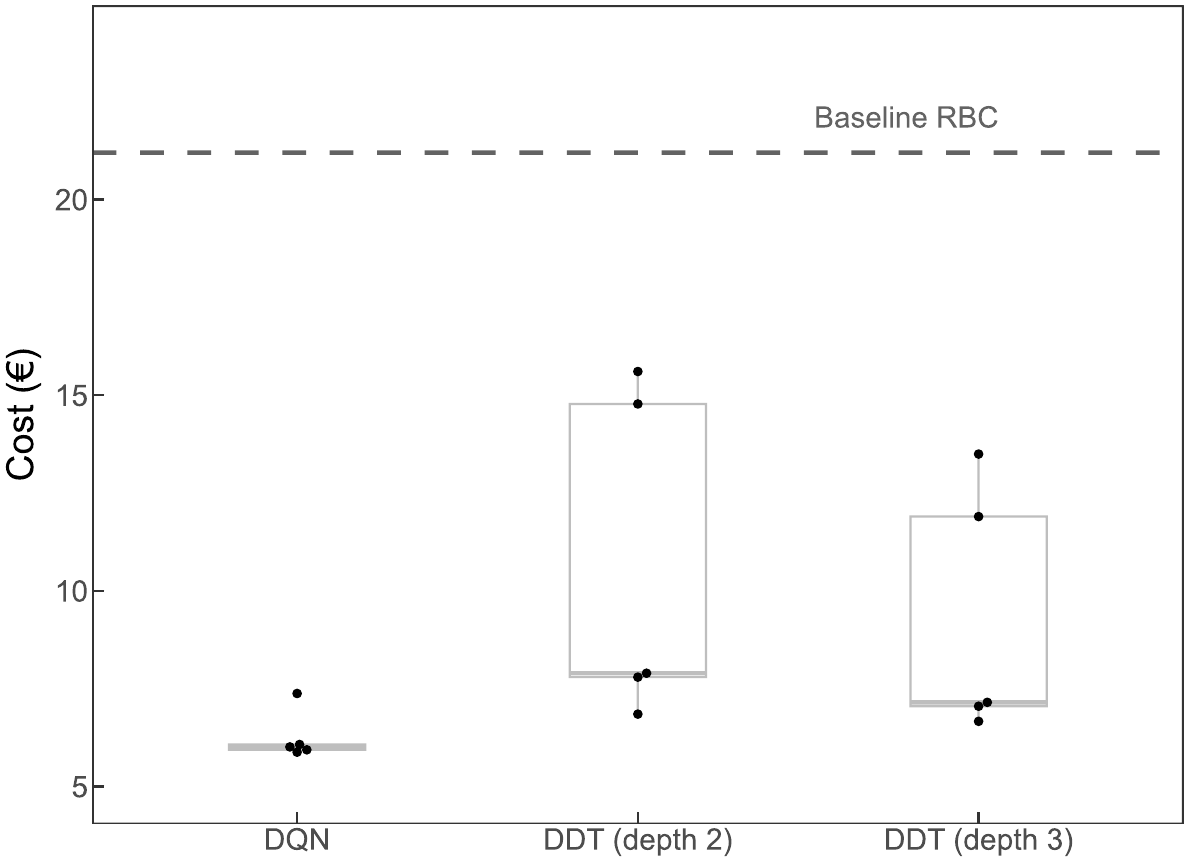}
         \caption{Artificial, square wave price profile}
         \label{fig:sq}
     \end{subfigure}
     \hfill
     \begin{subfigure}[b]{0.45\textwidth}
         \centering
         \includegraphics[width=\textwidth]{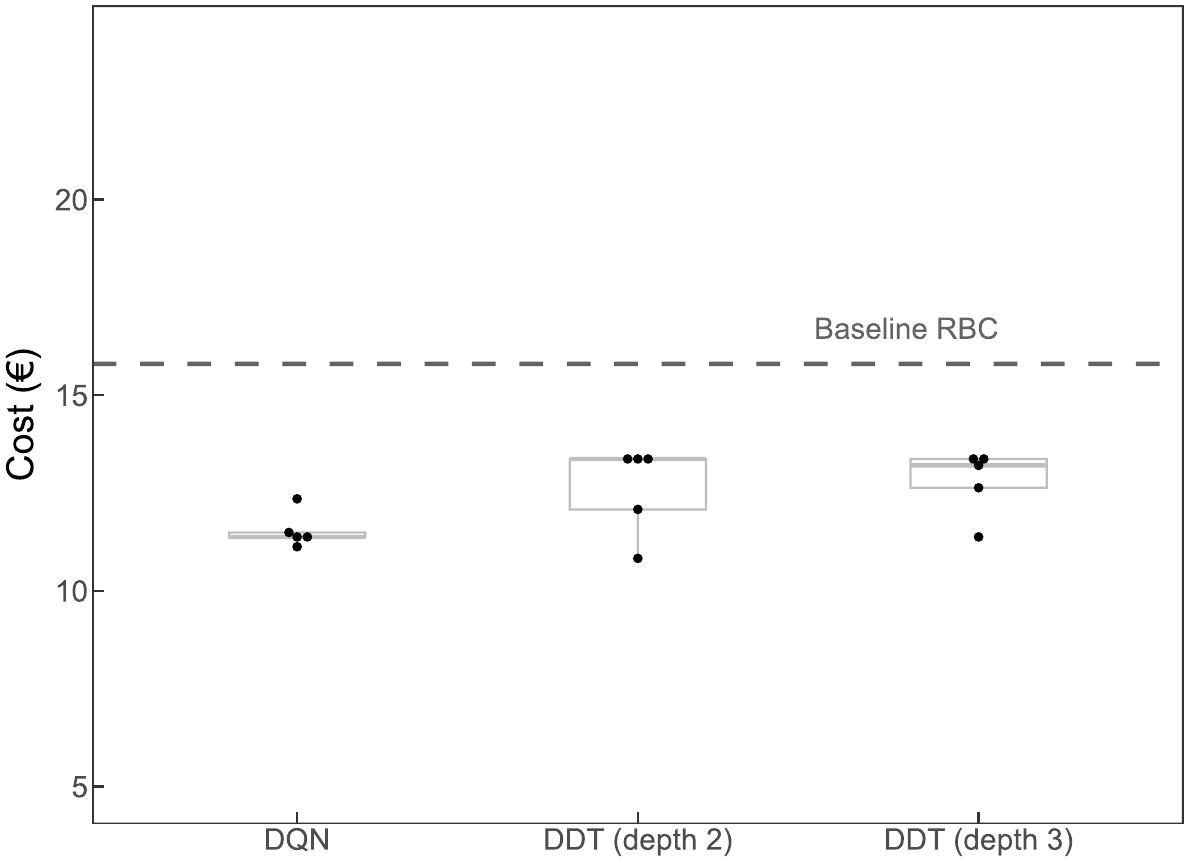}
         \caption{Real-world BELPEX price profile}
         \label{fig:belpex}
     \end{subfigure}     
\caption{Performance of DDT-based students as a HEMS on different price scenarios. The dots represent the actual performance of individual models and the box plots show the aggregate performance. The student agents are benchmarked using teacher agent ``DQN'' and a RBC.}
\label{fig:performance}
\end{figure}

The performance of our proposed approach using DDTs of depth 2 and 3 is presented in~\cref{fig:performance}. We note two key observations: 
\begin{enumerate*}[(i)]
    \item both DDT agents clearly outperform the baseline RBC controller;
    \item while the DQN-based teacher performs better than the DDTs, the performance difference (mean) is quite small ($\sim 5\%$). 
\end{enumerate*}
This indicates that our proposed approach can learn satisfactory control policies that outperform an RBC included with standard batteries. Additionally, the DDTs can mimic the teacher agents well and sustain minimal deterioration in performance. 

While the overall performance is satisfactory, \cref{fig:performance} indicates some (training) stability issues with the DDT-based controllers. This is particularly apparent in~\cref{fig:sq}, where both DDTs demonstrate a strong performance for 3 of the runs, while the other two instances do not fare as well. This problem can be attributed to the training process, where changes in `upstream' or hierarchically higher features could have a disproportionate impact on the output distributions. This needs to be investigated further and will be part of future work as discussed in~\cref{subsec:future}. 
\begin{figure}[t]
     \centering
     \begin{subfigure}[b]{0.47\textwidth}
         \centering
         \includegraphics[width=\textwidth]{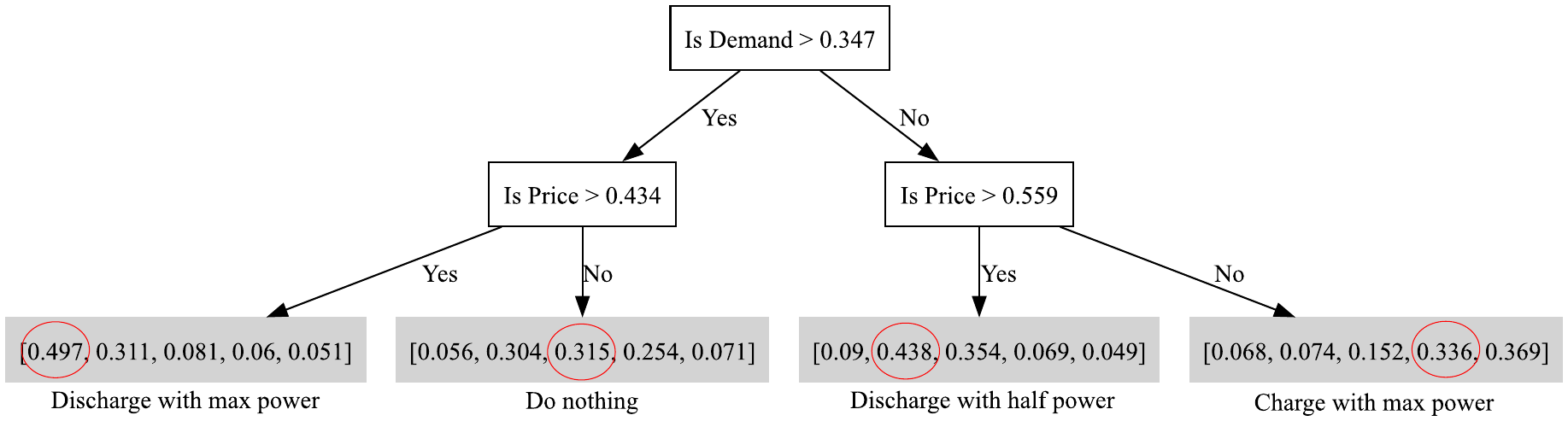}
         \caption{learned DDT for Square wave price scenario}
         \label{fig:sq_ddt_2}
     \end{subfigure}
     \hspace{1em}
     \begin{subfigure}[b]{0.47\textwidth}
         \centering
         \includegraphics[width=\textwidth]{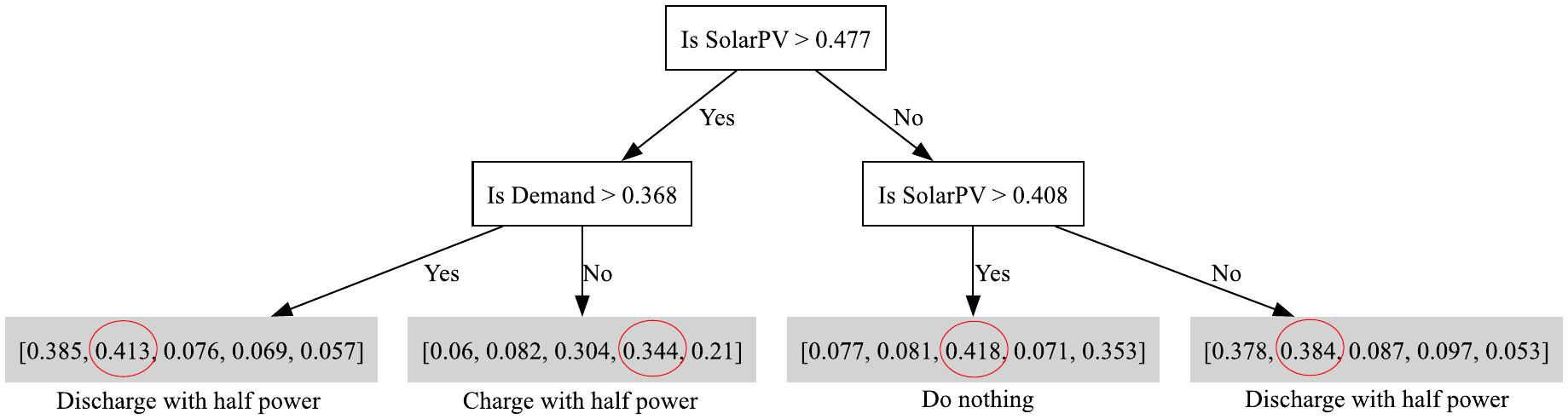}
         \caption{learned DDT for real-world price scenario}
         \label{fig:belpex_ddt_2}
     \end{subfigure}     
\caption{Visual representation of learned decision trees of depth 2 for both price scenarios. The decision nodes are depicted with unshaded boxes and contain the learned features and the threshold values. The leaf nodes are depicted by grey boxes and contain the learned distribution. The annotations highlight the actions related to each leaf node.}
\label{fig:learned_ddt}
\end{figure}

Furthermore, examples of learned DDTs of depth 2 are presented in~\cref{fig:learned_ddt}. Note that these DDTs are randomly initialized and over the course of training learn the feature selection~(e.g., choosing `demand' or `solar PV' as the feature for the first decision node) and the respective cut thresholds via gradient descent. We observed that both DDTs are straightforward to understand, easily `explaining' how the controller takes an action. Additionally, the actions taken are intuitive and follow human intuition -- e.g., in~\cref{fig:belpex_ddt_2}, the controller decides to take a charging action only when solar PV generation is high (greater than 0.47) while demand is low (less than 0.37). Likewise, in~\cref{fig:sq_ddt_2}, the controller discharges with maximum power when both price and demand are high while only discharging by half the power when price is high but demand is low, showing that the learned policy adjusts its decision based on the current as well as expected future demand. 

We conclude that the results presented in~\cref{fig:performance} and~\cref{fig:learned_ddt} validate the performance and explainability of our proposed DDT-based approach and show that the DDTs learn a simple, easy-to-explain policy and achieve satisfactory control performance.

\subsection{Explainability Comparison}
\label{subsec:exp_eval}
\begin{figure*}[t]
     \centering
     \begin{subfigure}[b]{0.48\textwidth}
         \centering
         \includegraphics[width=\textwidth]{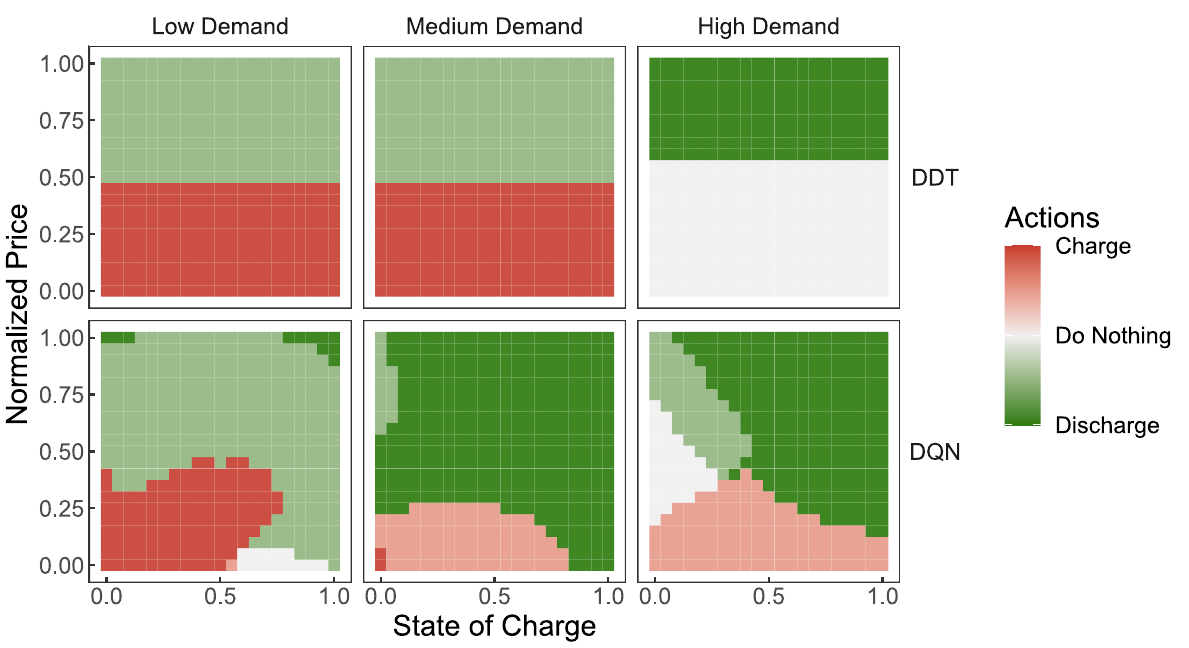}
         \caption{DDT of Depth 2}
         \label{fig:ddt_2}
     \end{subfigure}
     \hfill
     \begin{subfigure}[b]{0.48\textwidth}
         \centering
         \includegraphics[width=\textwidth]{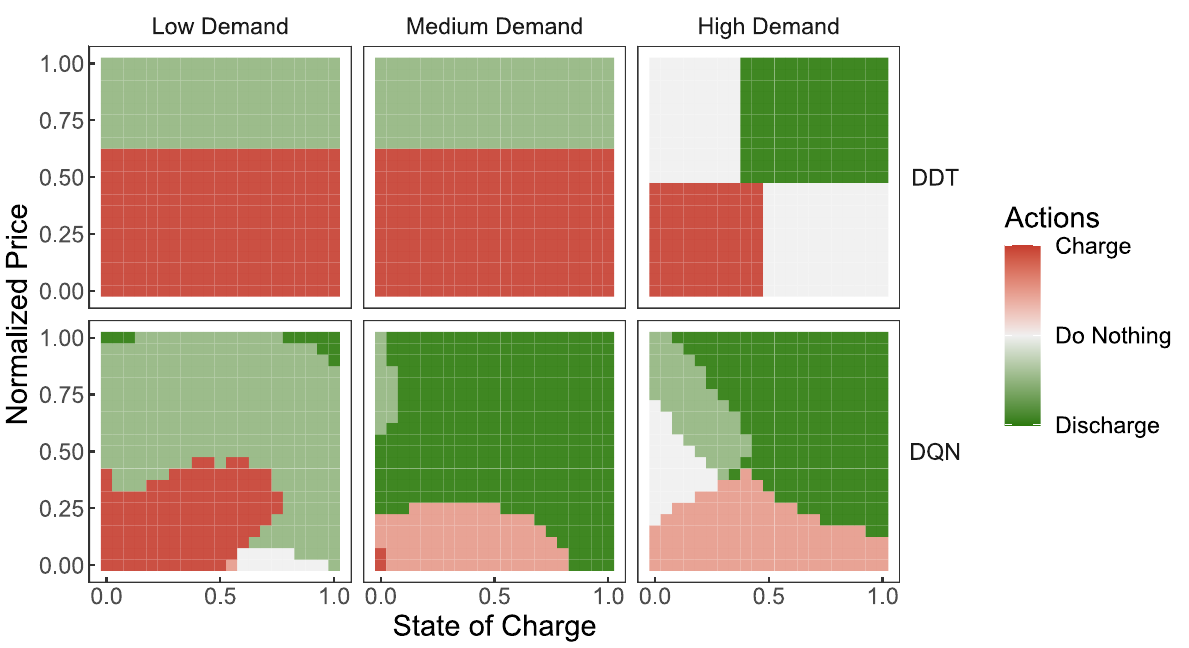}
         \caption{DDT of Depth 3}
         \label{fig:ddt_3}
     \end{subfigure}     
\caption{Visualizing the trained policy of DDT and DQN-based agent on a simplified HEMS scenario. The heatmaps show the actions chosen by the agents for different values of state-of-charge and price across different demand regions. The bottom row depicts the DQN policy and the top rows show the policy of our proposed DDT-based controllers}
\label{fig:explainability}
\end{figure*}

While~\cref{subsec:performance_eval} investigated the control performance of our method, we now examine the explainability of the obtained policies. As described in~\cref{subsec:scenarios}, we consider a reduced problem where a house without solar PV is exposed to an artificial square wave price profile. Despite being hypothetical, this scenario reduces the dimensionality of the state space~(now reduced to battery state-of-charge, price and non-flexible demand) and allows us to examine and compare the explainability of the learned DDT policy with that of the DQN policy. While the visual representation of the policy as shown in~\cref{fig:learned_ddt} is useful, we cannot directly compare it with the teacher policy~(which is a neural network). Consequently, we make use of policy heatmaps to visualize different control policies~\cite{qunat-xai}. \Cref{fig:explainability} illustrates such heatmaps comparing the teacher~(DQN) policy with depth 2 and depth 3 DDT policies. These heatmaps are generated by evaluating the controller's policy on all possible states (in a fixed subset of the state space) and provide an overview of how an agent would react for different states. 

Based on~\cref{fig:explainability}, we observe that the DDT heatmaps~(top rows of the figure) are consistent, straightforward, and can be easily decomposed into a few rules based on demand, price or state-of-charge. Contrary to this, the DQN-based policy is complex and often non-intuitive in terms of actions taken in specific regions. E.g., the DQN policy in the low-demand region prefers to discharge the battery even in the regions where the price is quite low (e.g., regions where price is less than 0.25 and the state of charge is greater than 0.75). Such behavior is counter-intuitive and difficult to understand even for experts, not to mention everyday homeowners (who will actually use such a system). This further highlights the increased explainability achieved using our proposed approach.

\subsection{Compute performance}
Besides explainability, the proposed DDT-based method is computationally light and easy to deploy on any edge device given that it reduces the control policy into a limited set of if-then-else rules. As a comparison, \cref{tab:compute} lists the number of parameters used and the storage footprint of the teacher agents and the distilled DDTs used in~\cref{subsec:performance_eval}. For DDTs, the number of parameters used during training are represented inside parenthesis along with the parameters used during inference. Unlike DQN which uses the same set of parameters during training and inference, DDTs require fewer parameters for inference -- e.g., at any decision node, the feature selection parameters can be reduced to a single parameter representing the selected feature. From this table, it can be clearly observed that the proposed DDTs have a significantly smaller compute footprint due to the reduced number of parameters, leading to trained models which are about 200 times smaller than the teacher DQN agents. To conclude, the comparison in~\cref{tab:compute} further underscores the potential for deploying such controllers in real-world scenarios.  

\begin{table}[t]
\centering
\caption{Comparison of DQN and DDTs based on computational metrics}
\begin{tabular}{c c c}
    \toprule
        Algorithm & Number of Parameters & Storage Size\\
    \midrule
        DQN (teacher agent) &  4.8k  & 22KB \\
    \midrule
        DDT -- depth 2 &  10 (38) & 4KB \\
    \midrule
        DDT -- depth 3 &  22 (82)  & 7KB \\
    \bottomrule
\end{tabular}    
\label{tab:compute}
\end{table}



\section{Conclusion}
\label{sec:conclusion}
Through this work, we introduced a novel method for obtaining explainable RL-based control policies using differentiable decision trees and policy distillation. The key idea of our work is to distill knowledge from a standard RL-based controller into a simple, easy-to-explain decision tree architecture by purely relying on data. For this, we use differentiable decision trees in a policy distillation setup, training the decision trees using a standard (pre-trained) RL-based controller and gradient descent. The policy distillation step allows extracting knowledge from an RL-based controller, while the differentiable decision tree architecture constrains the policy to be simple and explainable at all times. 

We validated our method on a battery-based home energy management problem and investigated the control performance and explainability of our proposed approach. As presented in~\cref{sec:results}, our proposed approach learns a control policy that performs comparable to the teacher DQN agent, while being simple~(i.e., $\sim 200$ times reduction in number of parameters) and easy-to-explain. Furthermore, the performance of our DDT-based controllers surpasses the performance of commonly observed RBC, performing $\sim 20-25\%$ better than the RBC. 

\subsection{Limitations and Future Work}
\label{subsec:future}
As discussed in~\cref{sec:intro}, the goal of this work was to introduce this novel method and highlight its potential for future applications in the energy domain. In support of this objective, we identify some limitations within the current work and outline areas for future investigations. 

The initial consideration pertains to the problem formulation discussed in~\cref{subsec:problem}, which we will further expand to include thermal models and joint optimization with comfort constraints. While the current problem mimics a real-world house in the present times, future application scenarios will require more elaborate HEMS that can optimize cost by leveraging flexibility from different sources including building thermal mass, batteries, and electric vehicles. To efficiently deal with such complex scenarios, our future work will explore two main aspects:
\begin{enumerate*}[(i)]
    \item extending the policy distillation set-up to multi-agent RL settings, where simple, shallow DDTs can be trained per flexibility asset; and
    \item domain knowledge induced feature engineering (using previous works such as~\cite{physnet}) to compress information and allow the use of shallow DDTs
\end{enumerate*}.
While large DDTs can be trained for such complex scenarios, we intend to focus on ``shallow'' DDTs that are intuitively easier to explain (or more explainable) as compared to ``deep'' trees.

Besides this, another limitation of our current approach is the occasional instability in the training process related to the DDTs. As noted in~\cref{sec:results}, this training instability could be attributed to the tree structure of the DDT with features hierarchically higher up in the tree significantly affecting the outputs. This needs to be investigated further to identify possible solutions to stabilize the learning process. This includes effective regularization strategies, warm starting, or constraining the decisions being learned. The latter seems particularly useful for DDTs of higher depth, where some decisions are conflicting or redundant~(as shown in~\cref{sec:add-ddt}). 

The third area that needs to be addressed further is the deployment of such an algorithm in real-world scenarios and performing a user trial to further validate the explainability of our method. While non-trivial, such a pilot study is needed to investigate the acceptance of such a HEMS as well as the challenges associated with maintaining such a system. This will further allow us to investigate more advanced approaches such as human-in-loop training and intervention strategies to maximize the decision tree architecture and develop a robust, data-driven controller that can be widely deployed across houses. 



\bibliographystyle{ACM-Reference-Format}
\bibliography{ref-list}



\appendix

\section{Hyperparameters}

\subsection{Home Energy Management Environment}
\label{subsec:battery_hp}
For the HEMS simulator described in~\cref{subsec:simulator}, we consider a nominal power rating~($P^{\text{agg}}_{t}$) of 4kW, prompting the agents to perform peak shaving operations to avoid exceeding this value. The agents can do this by controlling the battery~(modeled based on~\cref{subeq:bat_mod}), the details of which are tabulated in~\cref{tab:battery_hp}. The action space represents the charging or discharging signal that is given to the battery (and not the actual power). 
\begin{table}[t]
\centering
\caption{Parameters related to the Battery model used in the Home Energy Management Simulator}
\begin{tabular}{cc}
    \toprule
     Parameter                  & Value     \\
    \midrule
    Max Capacity                & 10 kWh     \\
    \midrule
    Max Power                   & 4 kW       \\
    \midrule
   Efficiency (round trip)      & 0.9       \\
   \midrule
   Action Space                 & $\{ -1, -0.5, 0, 0.5, 1\}$ \\
    \bottomrule
\end{tabular}
\label{tab:battery_hp}
\end{table}

\subsection{DQN-based Teacher Agents}
\label{subsec:dqn_hp}
The hyperparameters used for the DQN-based teacher agents are listed in~\cref{tab:dqn_hp}. Additionally, during the distillation process, the temperature $\tau$ is set to 0.03. 

\begin{table}[t]
\centering
\caption{Hyperparameters for $Q$-network (DQN)}
\begin{tabular}{cc}
    \toprule
     Parameter              & Value             \\
    \midrule
    Optimizer               & Adam              \\
    \midrule
    Learning Rate           & 0.001            \\
    \midrule
    Activation Function     & $\text{ReLU}$     \\
    \midrule
    Mini Batch Size         & 1000              \\
    \midrule
    Output Size             & 5 ($= |\mathbf{U}|$)       \\
    \midrule    
    Hidden Layers           & [64, 64]          \\
    \midrule
    Target Update~($\tau$)  & 0.1               \\
    \midrule
    Buffer Size             & 5000              \\
    \bottomrule
\end{tabular}
\label{tab:dqn_hp}
\end{table}



\subsection{Baseline Rule-based Controller (RBC)}
\label{subsec:rbc}
We compare the performance of our proposed DDTs with the corresponding teacher agents and a baseline RBC. This baseline is modeled based on built-in controllers that are available with commercially available batteries and are generally designed to maximize the self-consumption of solar PV. Such an RBC is modeled as:
\begin{equation}
    u_{i} =   \begin{cases}
                -1                                          \ \ &: P^{\text{agg}}_{t} \leq -P^{\text{max}}    \\
                1                                            \ \ &: P^{\text{agg}}_{t} \geq P^{\text{max}}    \\
                \frac{P^{\text{}}_{t}}{P^{\text{max}}}        \ \ &: \text{otherwise}
            \end{cases},
\label{eq:rbc}
\end{equation}
where $P^{\text{}}_{t} = P^{\text{con}}_{t} + P^{\text{pv}}_{t}$ represents the power required or left over after self-consumption.



\section{Additional Example of Learned Differentiable Decision Trees}
\label{sec:add-ddt}
\begin{figure*}[h]
  \centering
  \includegraphics[width=0.9\textwidth]{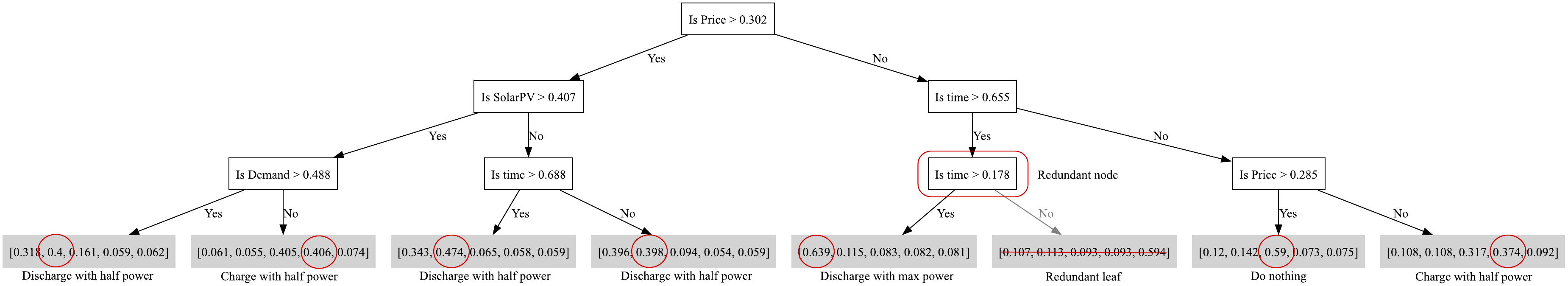}
  \caption{Example of a learned DDT for depth 3 for the real-world BELPEX price scenario.}
  \label{fig:ddt_3_learned}
\end{figure*}

\begin{figure*}[h]
    \centering
    \includegraphics[width=0.9\textwidth]{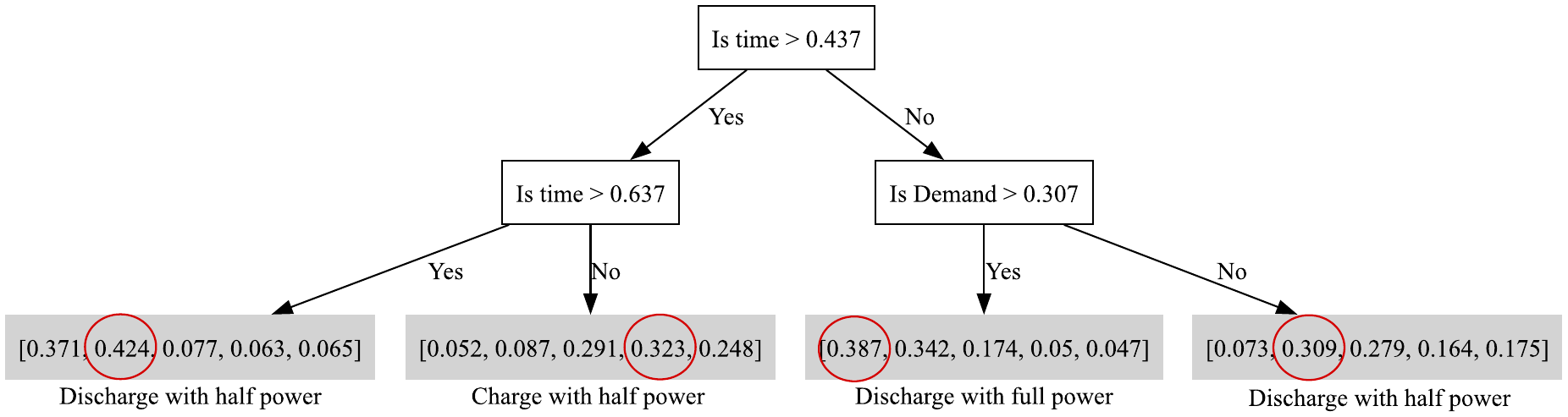}
  \caption{Example of a learned DDT for depth 2 for the real-world BELPEX price scenario. This particular policy learns an unusual set of rules, yet performs well,}
  \label{fig:ddt_2_weird}
\end{figure*}

This section provides some more examples related to the learned DDT policies. \Cref{fig:ddt_3_learned} depicts a depth 3 DDT student. The shown policy (decision tree) takes intuitive actions, e.g., charging the battery when the price is low or when solar PV is high. However, as compared to the depth 2 DDT (shown in~\cref{fig:belpex_ddt_2}), this policy is a bit more difficult to understand. Further, as annotated in the figure, some decision nodes learn redundant rules, which may lead to some training instability. Besides this, for some instances, the DDTs may learn rules that are not obvious or human intuitive, an example of such a scenario is presented in~\cref{fig:ddt_2_weird}. As shown in the figure, the policy bases its decisions on time as a feature, possibly exploiting the underlying correlation between time, price, and solar PV generation. Hence, while the rules learned are unusual, this policy still shows good performance. Such unusual rules are related to high correlation between various features and to improve the rules~(making them more intuitive) we need to carefully select the relevant features from the perspective of information content and explainability.  



\end{document}